# A case study of multi-modal, multi-institutional data management for the combinatorial materials science community

Sarah I. Allec[1], Eric S. Muckley[1,7], Nathan S. Johnson[2,8], Christopher K. H. Borg[1,9], Dylan J. Kirsch[3,4], Joshua Martin[4], Rohit Pant[3], Ichiro Takeuchi[3], Andrew S. Lee[5,6], James E. Saal[1*], Logan Ward[6], Apurva Mehta[2]

[1]*Citrine Informatics, Redwood City, 94063, CA, USA*

[2]*SLAC National Accelerator Laboratory, Menlo Park, 94025, CA, USA*

[3]*Department of Materials Science & Engineering, University of Maryland, College Park, 20742, MD, USA*

[4] *Material Measurement Laboratory, National Institute of Standards and Technology, Gaithersburg, 20899, MD, USA*

[5]*Department of Materials Science & Engineering, Northwestern University, Evanston, 60208, IL, USA*

[6]*Argonne National Laboratory, Lemont, 60439, IL, USA*

[7]*Currently at 11:59, Sacramento, 95822, CA, USA*

[8]*Currently at ZEISS Microscopy, Dublin, 94568, CA, USA*

[9]*Currently at NobleAI, San Francisco, 94104, CA, USA*

*Corresponding author. E-mail: jsaal@citrine.io (ORCID: 0000-0003-0935-158X)

Contributing authors: sallec@citrine.io (ORCID: 0000-0002-1101-3160), eric.muckley@elevenfiftynine.com (ORCID: 0000-0001-7114-5424), nathan.johnson@zeiss.com (ORCID: 0000-0002-6565-7236), chris@cborg.dev (ORCID: 0000-0002-2641-3319), dylan.kirsch@nist.gov (ORCID: 0000-0001-5132-3457), joshua.martin@nist.gov (ORCID: 0000-0003-2214-0955), rohitp@umd.edu (ORCID: 0000-0002-4591-5752), takeuchi@umd.edu (ORCID: 0000-0003-2625-0553), andrew.lee@anl.gov (ORCID: 0000-0001-5301-4295), lward@anl.gov (ORCID: 0000-0002-1323-5939), mehta@slac.stanford.edu (ORCID: 0000-0003-0870-6932)

# Abstract

Although the convergence of high-performance computing, automation, and machine learning has significantly altered the materials design timeline, transformative advances in functional materials and acceleration of their design will require addressing the deficiencies that currently exist in materials informatics, particularly a lack of standardized experimental data management. The challenges associated with experimental data management are especially true for combinatorial materials science, where advancements in automation of experimental workflows have produced datasets that are often too large and too complex for human reasoning. The data management challenge is further compounded by the multi-modal and multi-institutional nature of these datasets, as they tend to be distributed across multiple institutions and can vary substantially in format, size, and content. Furthermore, modern materials engineering requires the tuning of not only composition but also of phase and microstructure to elucidate processing-structure-property-performance relationships. To adequately map a materials design space from such datasets, an ideal materials data infrastructure would contain data and metadata describing *i*) synthesis and processing conditions, *ii*) characterization results, and *iii*) property and performance measurements. Here, we present a case study for the low-barrier development of such a dashboard that enables standardized organization, analysis, and visualization of a large data lake consisting of combinatorial datasets of synthesis and processing conditions, X-ray diffraction patterns, and materials property measurements generated at several different institutions. While this dashboard was developed specifically for data-driven thermoelectric materials discovery, we envision the adaptation of this prototype to other materials applications, and, more ambitiously, future integration into an all-encompassing materials data management infrastructure.

# Introduction

Responding to the global energy crisis heightened by the COVID-19 pandemic [1], as well as facing long-standing global challenges such as climate change, will require the development of new, transformative materials on an accelerated timescale. While typical materials discovery and design occur on timelines measured in decades [2], today's need for renewable energy is urgent, as fuel prices, poverty, and geopolitical turmoil increase in response to fuel shortages [1]. While the design (compositional and processing parameter) space of all possible materials is practically unlimited, most of this space is unexplored and the discovery rate of transformative materials is relatively slow. Although the convergence of high-performance computing (HPC), automation, and machine learning (ML) has significantly altered this timeline [3–6], transformative advances in functional materials and acceleration of their design will require addressing the deficiencies that currently exist in materials informatics, particularly a lack of centralized, standardized data management.

The Open Science movement [7–10] has driven several significant efforts in data management generally (e.g., FAIR (Findable, Accessible, Interoperable, Reusable) data principles [11]), particularly through its influence on the development of data-related scientific policy now enforced by journals [12, 13] and funding agencies, such as the U.S. Department of Energy's Public Access Plan [14]. For materials datasets like the Materials Project [15] and the Open Quantum Materials Database (OQMD) [16, 17], where all of the data are produced computationally according to a standardized methodology and format, the creation of a data infrastructure that follows FAIR principles is more straightforward than for the case of experimental data, which can vary significantly based on how, where, and when they were produced [2]. These challenges associated with experimental data management are particularly true in the case of combinatorial materials science, where advancements in automation of experimental workflows have produced datasets that are often too large and too complex for human reasoning, necessitating the use of data science techniques for knowledge extraction and interpretation. This challenge is further compounded by the multi-modal and multi-institutional nature of these datasets, as they tend to be distributed across multiple institutions and can vary substantially in format, size, and content. Furthermore, many institutions host a breadth of legacy data that is essentially untouched by modern materials informatics techniques, sitting in storage and yet to be explored.

Another additional dimension of complexity is the need to record the entire materials life cycle in order to elucidate processing-structure-property-performance (PSPP) relationships, a feat that computational databases currently do not achieve. Modern materials engineering requires the tuning of not only composition, but also of phase and microstructure, all of which influence PSPP relationships. By design, combinatorial thin-film libraries provide complete determination of compositional, structural, and materials properties, offering a wealth of information if combined and organized in a meaningful way [18–20]. To adequately map a materials design space (MDS) from such datasets, an ideal materials data infrastructure would contain data and metadata describing i) synthesis and processing conditions, ii) characterization results, and iii) property and performance measurements. As all of these data types are usually stored in different files and possibly hosted in different locations, the barrier to organizing vast amounts of such fractured data is high for individual research groups. A tool for the automated organization, analysis, and visualization of multi-modal data has the potential to rapidly generate PSPP relationships, as well as uncover new insights by combining datasets across institutions, projects, and time into larger, more comprehensive databases. This could be especially transformative for high-throughput experimentation (HTE), where data analysis is often the rate-limiting process. [21]

Here, we present a case study for the development of a web-based dashboard that enables standardized organization, analysis, and visualization of a large data lake consisting of combinatorial datasets of synthesis and processing conditions, X-ray diffraction (XRD) patterns, and materials property measurements produced at several different institutions. At the core of this work is data standardization, persistence, and accessibility across institutions. While desktop packages [22, 23] exist for the analysis and visualization of combinatorial materials data and there are web-based platforms for displaying publication-ready HTE data [24], we require a tool which combines the ability for scientists to interact with data while also providing a secure route for sharing data before larger release. Furthermore, the dashboard directly integrates with Globus[1] [25, 26], a tool already widely used by the research community, for file storage and user authentication. A web-based dashboard removes any need for the user to store or download data locally, reduces the learning curve for usage, and provides data security by default. Throughout the following sections, we document the challenges faced and solutions developed during the evolution of the dashboard, as well as recommendations for the low-barrier development and incentivization of multi-modal, multi-institutional data management within the materials science community at large. While this dashboard was developed specifically for a project on data-driven thermoelectric materials discovery, we envision the adaptation of this prototype to other materials applications, and, more ambitiously, future integration into an all-encompassing materials data management infrastructure.

---

[1] Certain commercial equipment, instruments, software, or materials are identified in this document. Such identification does not imply recommendation or endorsement by the National Institute of Standards and Technology, nor does it imply that the materials or equipment identified are necessarily the best available for the purpose.

## Methods

**Project description**

The data portal described here was developed during a ≈4.5 year multi-institutional project on thermoelectric materials discovery, ThermoElectric Compositionally Complex Alloys (TECCA). The aim of this project was to accelerate the discovery of sustainable and inexpensive high-performing thermoelectric (TE) materials by combining artificial intelligence and high-throughput computations and experiments. High TE performance requires a delicate balance among competing properties, requiring a fine tuning of the composition (through solid-solution and doping) and of processing to achieve an optimal hierarchically tailored microstructure [27]. The needed additional optimization greatly enlarges the composition-structure-microstructure combinatorial search space, and without accurate guidance device-quality materials are very difficult to find, as well as to synthesize. On the experimental side of this project, multiple institutions were involved in manufacturing samples, with SLAC and NIST performing the diffraction and thermoelectric measurements, respectively, as displayed in **Table 1**. The data produced at each institution, most of which was produced during the first two to three years, were aggregated into the multi-modal, multi-institutional dataset managed by the dashboard. During the third year of the project, the limitations of Globus for collaborative and persistent data analysis and visualization initiated development of a project-specific data dashboard, which was developed and used during the last two years of the project. While the exact number of personnel fluctuated throughout the project, the team generally consisted of about ten experimental researchers and five theoretical researchers. Visualizations and resulting insights gained from the dashboard have been used in one PhD thesis to date.

**Table 1.** Types of measurements performed by each institution in the TECCA collaboration. "Sample" refers to manufacturing of samples.

| Institution | Sample | Composition | Diffraction | Thermoelectric |
|---|---|---|---|---|
| Northwestern University | x | | | |
| University of Maryland | x | x | | |
| SLAC | x | x | x | |
| NIST | x | x | | x |

**Data portal**

The development of a data portal was sparked by the desire to increase collaboration amongst a multi-institutional experimental team. The primary goal was to create the ability to upload data to a location accessible to the entire team and to organize (i.e., index) the data for easy accessibility and usability. In addition, it was desirable that the data be available for processing, analysis, and visualization directly in the dashboard, without needing to download the data locally. To best meet these needs, it was decided that the dashboard would need a graphical user interface (GUI) accessible via a web browser, with a Representational State Transfer (REST) API [28] for programmatic access to the database and a backend server for data processing (e.g., processing, indexing, searching, and aggregating uploaded data). The GUI is written in the scalable JavaScript framework via Svelte [29], a free and open-source frontend framework, and the backend utilizes Flask, a micro web framework written in Python [30]. The accompanying RESTful API is written in Python. All data are stored in the Globus cloud file storage system [25, 26] on a single endpoint located at the Argonne Leadership Computing Facility (ALCF), although future developments could use multiple endpoints for federated database management. Lastly, the data portal design follows the established Globus Modern Research Data Portal Design Pattern [31] and is hosted on a Heroku [32] server.

A schematic of data flow into and within the data portal is shown in **Figure 1**. Flow into the platform begins when a team member uploads data files to Globus, which are then automatically processed, organized, and aggregated into the database via custom processing routines that utilize the Globus Python SDK in the backend. Due to the multi-modal nature of the data in this project, as well as the lack of standardization in file formatting and content, several ingestion scripts were written for the different file types in order to standardize formatting and naming conventions. Custom indexing scripts then organize and aggregate the standardized data across different experiments performed throughout the project (details regarding this standardization are given in the Results section). Finally, data flows into the frontend, which is accessed by the user via a web browser. On the webpage, the user can view the aggregated data in a tabular format, search and filter the data, and visualize the data in several different plot types. Each of these features is discussed in more depth in the following sections.

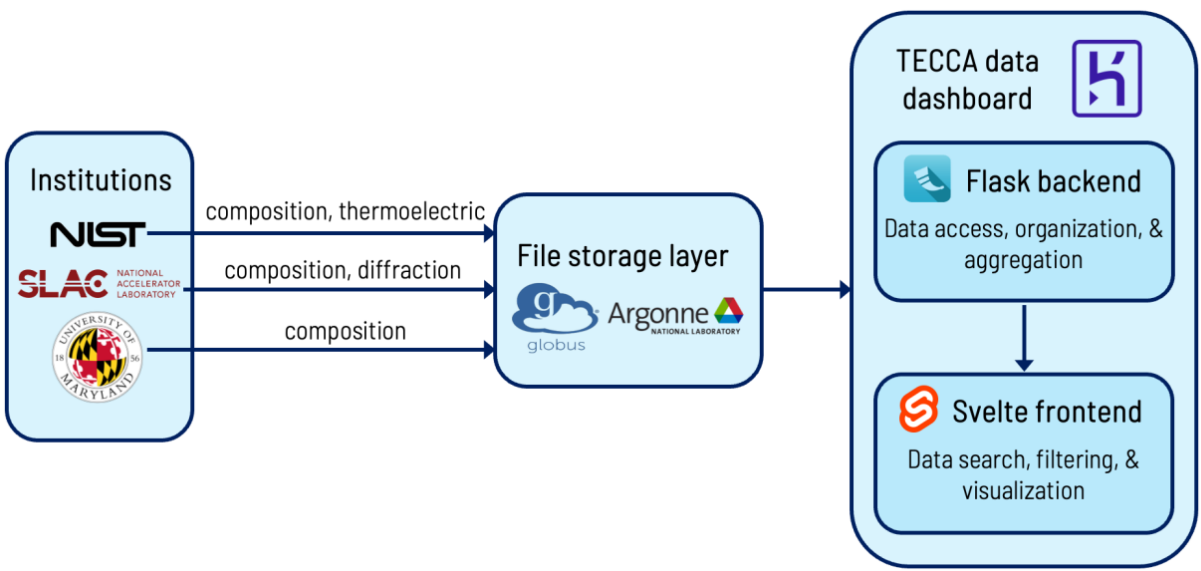

**Figure 1.** Schematic of data flow into and within the web-based data portal: First, data from the experimental team is uploaded to the Globus endpoint located at ALCF, which is then processed via the backend and accessible by the users via the frontend, all of which is hosted on a Heroku server.

## Results

Here, we provide the reasoning behind our design choices, as well as further details regarding each dashboard feature. We also identified several barriers to the development of complex data infrastructure within the materials science research community and how our design choices overcome these barriers.

### Data storage and user authentication

Apart from data standardization, the primary barriers to developing infrastructure for multi-institutional materials data sharing and management are *i*) data security and *ii*) large storage capabilities. Some of the concerns around data security are cultural, particularly those regarding pre-publication data sharing within a project, but another aspect is simply that material scientists are not aware of the best practices for enabling robust security and user authentication. Even if these issues are resolved, password fatigue alone may be a deterrent to incentivization for the wide adoption of a new tool. With respect to file storage, most research groups have access to a limited amount of storage in up to a few distinct locations, but may not have the storage or transfer capabilities to move and host all data across a project, not to mention legacy data or relevant data from other sources (*e.g.*, open source databases). In our case, the size of the XRD dataset alone was nearly 0.5 TB. These data needed to be accessed by multiple institutions and combined with other datasets in order to elucidate PSPP relationships; storing and transferring such large amounts of data amongst institutions would have been costly in term of both time and money. Lastly, another associated barrier is that cloud-based large file storage, security, and website hosting are expensive and require expertise to develop, necessitating either the hiring of experts or the development of such skills within the group. Consequently, the incentives for building data infrastructure for multi-institutional, multi-modal data often do not outweigh the perceived learning curve and cost.

In addition to these challenges, multi-institutional data are hard to aggregate in one place because it has been generated at geographically-distributed locations, and there tends to be technical and cultural overhead associated with moving it. We addressed this and the aforementioned challenges by utilizing Globus as the data storage layer for the dashboard. Globus was chosen because i) it is already widely used by the scientific research community and ii) it offers several data management capabilities, including efficient high-speed data transfer, interoperability with existing infrastructure via the Globus Python SDK, and native support for automated workflows via Globus Flows. Globus also provides native data security features, including encryption, fine-grained user access controls, and OAuth2 [33] and OpenID Connect standards [34–36] for user authentication.

## Data organization and aggregation

The various datasets incorporated into the dashboard contained a variety of file types, including csv files, Microsoft Excel files, image files, and text files. These different file types contain different types of data, such as composition, XRD patterns, and material properties, as well as different shapes and sizes of data – encompassing a truly multi-modal dataset. In addition, there were inconsistencies in naming conventions and file formatting, which required tedious handling when merging the data in these files. This lack of standardization makes data aggregation particularly fragile, as any discrepancy in format or naming can lead to data corruption. Another dimension of complexity arises from the combinatorial aspect of these datasets, where each wafer contains multiple samples (i.e., spots) creating a gradient in composition and properties. For example, amongst all file types, there were several different column names for designating the wafer spot for each data point (see **Table 2, column 1**) and different keys in the XRD files (**Table 2, column 2**). For standardizing wafer spot names and XRD keys, as well as typos, standard names were selected and corresponding columns were renamed (*e.g.,* columns with any of the possible wafer spot names in **Table 2, column 1** were renamed as "Spot"). All aggregation was done with the merging methods of the Python pandas package.

**Table 2.** Variation in column names for multi-modal dataset, necessitating standardization across multiple file types and origins.

| Possible wafer spot names | Possible XRD keys |
|---|---|
| Spot # | _x_ |
| # | _x2_ |
| No. | _xrd_ |
| Pad No. | resolution |
| Pad# | _1d |

Because of these factors, we wrote custom file ingestion scripts for standardizing file formatting and naming conventions and custom indexing scripts for generating large indexes of all the files and their associated data. These enabled aggregation of different types of data for the same material, as well as project-wide searching, sorting, and filtering of experimental data, which would be impossible in a traditional data lake consisting of different file and data types. A screenshot of the aggregated data for a selected wafer is shown in **Figure 2**. As a result of this aggregation, all of a wafer's data can be viewed by simply clicking on its name in the left-hand menu, resulting in the tabular display of all aggregated data for that wafer. Each row contains all of the measured property and compositional data for each spot on the wafer. Furthermore, each column can be sorted by simply clicking on the column name (e.g., "Pt Corrected Seebeck Coefficient (uV/K)") as well as filtered by searching for different values in the search box under the column name. We emphasize that the columns in this table originated from different files produced at different institutions and are stored in different folders on Globus. Furthermore, not all samples contained all data types – some only contain compositional data while others contain electrical properties and XRD spectra.

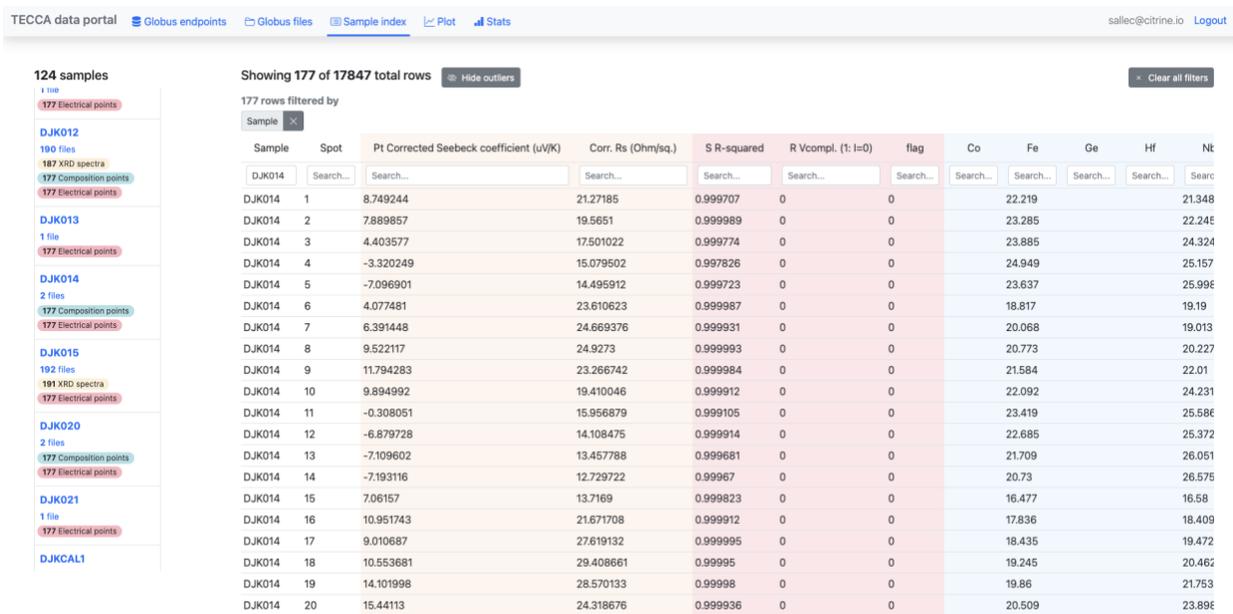

**Figure 2.** Screenshot of the aggregated dataset in tabular form, with wafer DJK014 selected. For each wafer, all available XRD, electrical, and compositional data is aggregated into a single table, where each row represents a different spot on the wafer.

## Data visualization

While data aggregation and organization of this complex multi-modal, multi-institutional data lake is a feat in and of itself, arguably the most impactful and scientifically useful features of the dashboard are its visualization capabilities, coupled with its capabilities for data persistence. There are four types of plots available in the dashboard: i) Ternary plots (ternary plots colored by a property or elemental fraction, **Figure 3**), ii) wafer plots (visualization of a property or elemental fraction over all spots on a wafer, **Figure 4**), iii) scatter plots (visual comparison of properties and/or elemental fractions of selected wafers, **Figure 4**), and iv) XRD plots (XRD spectra visualizations over different spots on a single or multiple wafers, **Figure 5**). All of these plots can be generated simply by selecting the plot type from the "New plot" button at the top of the plotting page (cf. **Figures 3 to 5**). Lastly, plot configurations can be saved via the "Save plot configuration" button to be shared and re-used via the "Open plot configuration" button, enabling data persistence. The plots currently available on the dashboard are those that are relevant to the data available and this particular project's goals; however, additional plot types can be added easily due to the modularity of the source code. We see this as a "plug-and-play" feature that can be utilized by materials scientists across diverse fields, as long as a script is available for generating the desired plots.

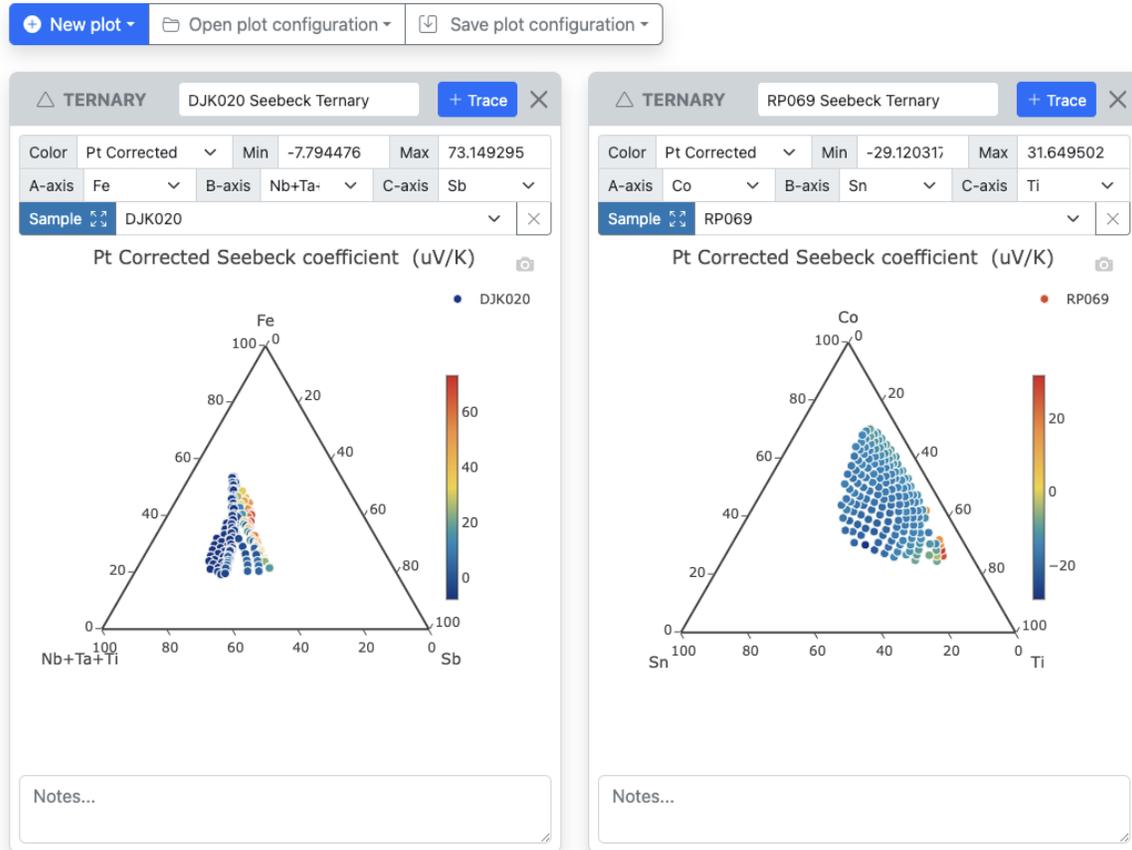

**Figure 3.** *Sample ternary plots for two wafers (DJK020 on the left, RP069 on the right) with each point in the ternary colored by Seebeck coefficient.*

The most valuable benefit of these visualization capabilities is the ability to compare multiple data types across multiple samples simultaneously. The nature of the collaboration (being both multi-institution and multi-modal) meant that data aggregation and analysis was often compartmentalized: one group focused on thermoelectric property measurement, another on sample synthesis, another on X-ray diffraction, etc. Originally, the dashboard reflected this: all plots were generated on each individual sample's page when you clicked on the sample name. However, this prevented the comparison of different wafers, leading to the development of a separate plotting page where different wafers and spots on wafers could be selected and compared. Prior to this development, comparing trends across both multiple samples and multiple data types was difficult without consistent coordination between teams. In a multi-modal characterization study, researchers often ask questions like "How do the trends in systems X, Y and Z compare to the trends in systems A, B, and C?" Each individual system alone can contain hundreds of data points for a single data modality, and concatenating hundreds of data points across systems becomes burdensome. Instead, the dashboard enabled a workflow where plots of data points across dozens of samples and multiple modalities could be generated quickly, efficiently, and consistently - researchers were able to make and share plots on-the-fly during meetings. This greatly increased collaboration, as ideas or questions about the aggregated data could be plotted and interrogated immediately.

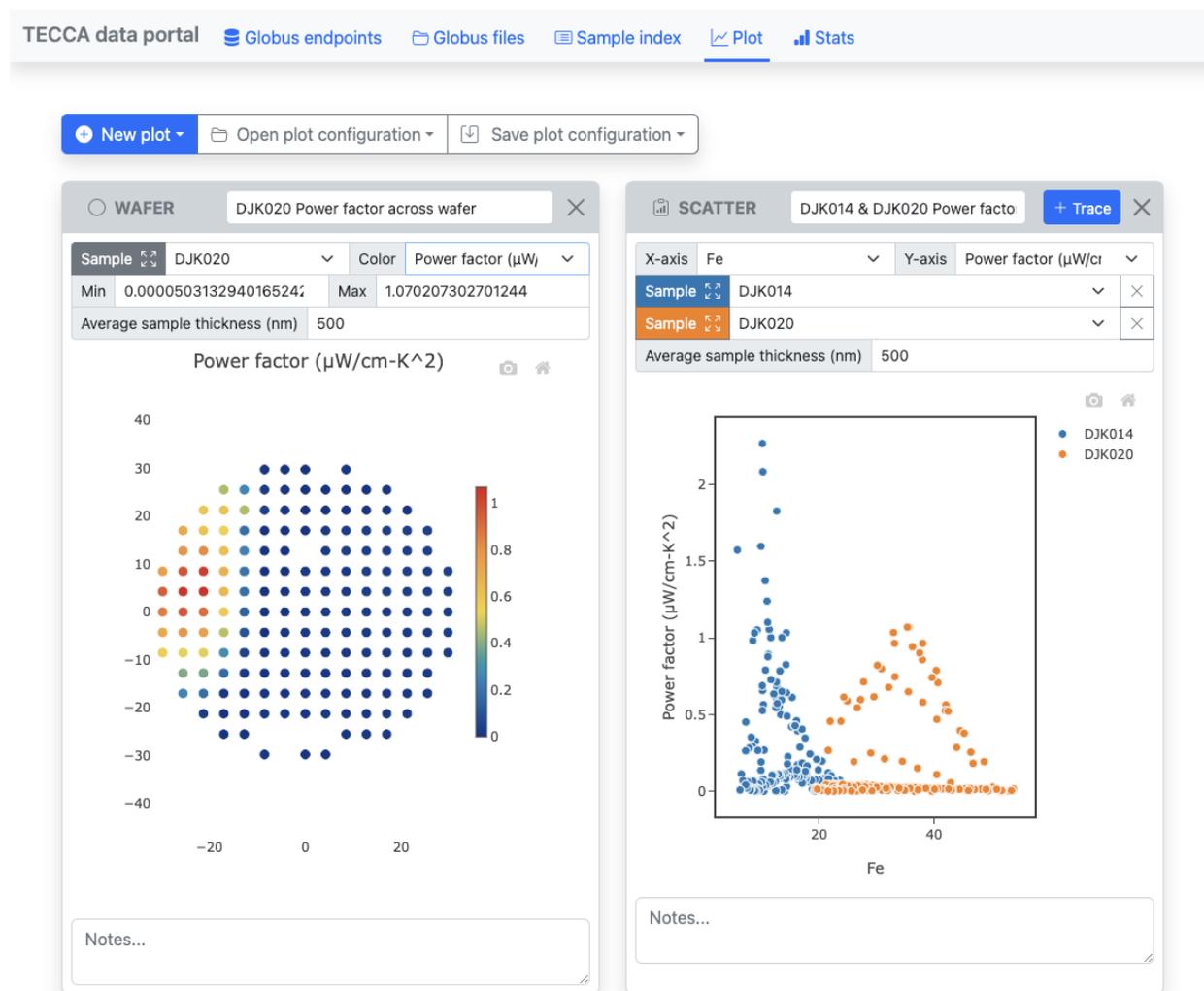

**Figure 4.** Sample (*left*) wafer plot with each spot on the wafer colored by power factor and (*right*) scatter plot of power factor vs. Fe content for two wafers (DJK014 and DJK020).

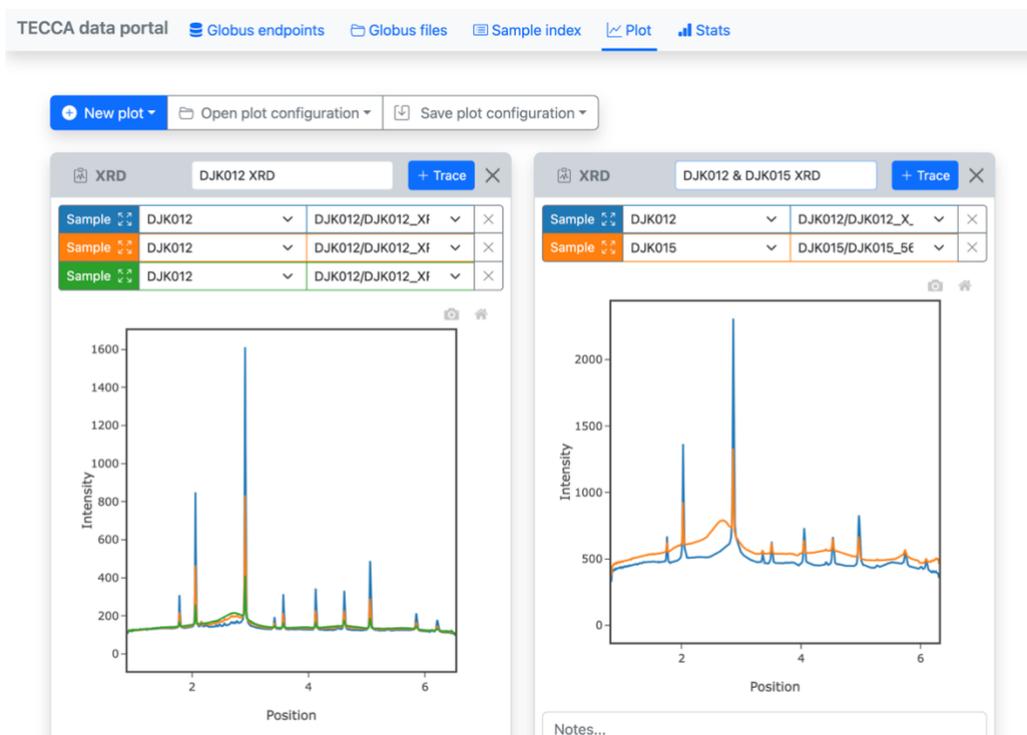

**Figure 5.** Sample XRD plots comparing the XRD patterns of (*left*) different spots on a single wafer (DJK012) and (*right*) specific spots on different wafers (DJK012 and DJK015).

# Discussion

Since the inception of the Materials Genome Initiative (MGI) [37], the utilization of data-driven methods within materials research has grown significantly and has permanently changed the nature of materials discovery and design. At the heart of the MGI is knowledge sharing, which is realized through the Materials Innovation Infrastructure (MII) – a framework of integrated advanced modeling, computational and experimental tools, and quantitative data. In light of the advances made in ML and automation for materials science during its first decade, in 2021 the MGI updated its strategic plan [38], with two goals particularly relevant to this work: i) the unification of the MII and ii) harnessing the power of materials data. While FAIR databases and associated infrastructures exist for computational materials data, there are far fewer instances of these for experimental materials data. Often, these tools are siloed within individual research groups or are not in widespread use due to a lack of awareness and/or incentives within the materials research community. In order to meet the new goals of the MGI, experimental data and associated tools need to be shared efficiently and effectively. Furthermore, unification of the MII requires some level of standardization of the individual tools to ensure that they are valuable, accessible, and easy to use. Toward this end, in the following sections, we describe the lessons we learned throughout the evolution of the portal, as well as future possibilities for materials data infrastructure.

## Lessons learned and recommendations

Here, we discuss the lessons we learned through the development and evolution of the data portal, as well as recommendations to the community in an effort to ease and quicken the development of similar tools and to provide some standardization across individual tools.

**Data standardization**

The major challenges faced during the building of the dashboard were related to a lack of data standardization. Some standardization conflicts were controllable, while others were due to the utilization of different experimental instruments, which varied in file format and naming conventions. In FAIR language, these individual datasets were not interoperable, and we imagine this to be the case for most multi-institutional datasets and legacy datasets, and is arguably the most challenging barrier to aggregating data across institutions and time. Our solution was to write many custom file ingestor scripts, with the standardization of naming conventions hard-coded. This solution works for a single project or for legacy data, and is easy to implement in Python with the pandas package, but is not general to the broader community, as the developer would need to communicate with all contributors in order to understand the various file formats and naming conventions used in each experimental set-up. It also requires significant maintenance overhead, as a developer would need to write a new ingestor script any time a new file format or naming conflict is encountered. However, hard-coding in Python is a low barrier method for establishing standards for a single project or legacy data, and can get the ball rolling in terms of creating an initial general infrastructure for different types of experiments and materials, as discussed in the Future Work section below.

With respect to standardization issues that are controllable (i.e., are chosen by researchers, not machines), these difficulties can be greatly alleviated by simply identifying and communicating standardized file formats and naming conventions at the beginning of a project. For handling standardization issues due to different experimental instruments, one potential solution would be to use natural language processing (NLP) to infer the structure of the data from different file formats and to learn what material property is associated with a given column name. Such an approach would both eliminate the need for multiple ingestor scripts and reduce maintenance overhead. While the training of such models still requires large amounts of data, we believe that the return on investment for these kinds of models will be high and that they are worth the initial effort of data curation and model training. Another potential solution is the designation and use of globally unique, persistent identifiers (PIDs) for scientific instruments, which would enable linking of data to the instrument that produced it and the storing of metadata regarding the instrument, as recommended by the Research Data Alliance Persistent Identification of Instruments Working Group. [39] Granted sufficient FAIR metadata is publicly available, a database of instruments widely used for a given type of experiment containing metadata regarding naming conventions and file formats could alleviate most data standardization issues. As this information can be proprietary, another solution could be a requirement by journals that researchers ensure that their data formatting adheres to standard formatting and naming conventions.

**Software development**

From the development side, our biggest takeaway was to not reinvent the wheel. Because this dashboard was built by and for materials scientists, it was important to use simple, commonly-used technologies that could be maintained by the team, as they did not have significant software development experience. As mentioned in previous sections, Globus provides established, trusted solutions for file storage, data security, and user authentication. Furthermore, Globus has developed a design pattern specifically for research portals, based on identified best practices for ensuring convenient, high-speed, secure access to large amounts of data via web-based portals. They also provide sample code skeletons that researchers can adapt to their needs, as was done in this work. Many scientific researchers are familiar with Globus and already use it in their own work, thus reducing the barriers to using a new tool, such as creating a new account (i.e., password fatigue), the learning curve associated with using a new tool, and implementing it in existing research workflows.

Because of this, a Python backend was chosen because most materials scientists are now familiar with Python for data analytics and plotting. Within Python, we selected Flask [30] - a lightweight, bare bones approach to backend. On the frontend, we followed a similar strategy: we used Sveltekit [29], which is simple and easy to run for non-developers. Furthermore, we stuck with the most common open-source library, Bootstrap [41], for styling the interface, and we used the most common set of icons for the interface, Bootstrap icons [42]. Using this set of standard, simple, and mature technologies was important for ensuring maintainability and feature development speed.

**Communication**

Key to the success of this work was extensive communication between the developer and the experimental team, not only regarding the technical development of the dashboard, but also to incentivize its usage: Weekly meetings were held to coordinate who was contributing which data and how they expected it to be displayed, visualized, and searched. Because of this frequent communication and feedback, the dashboard evolved continuously according to the needs of the experimental researchers. This not only informed feature development, but also ensured the accuracy and scientific relevancy of how the data is displayed in the dashboard and of the plotting features. These meetings also helped establish appropriate standardization for file formatting and naming conventions. Altogether, frequent communication between the developer and experimental team ensured that the domain experts in the field dictated the decisions made regarding feature development and data standardization.

**Competition with personal workflows**

A major motivation for the creation of the dashboard, as previously mentioned, was that collaborators were having difficulty comparing data cross modalities and samples due to the size, nature, and location of the datasets (i.e., data were spread across many institutions). The dashboard was introduced about three years into a 4.5 year collaboration. By the time the dashboard was introduced, each collaborator and institution had already developed their own internal methods for accessing and plotting data. Some users were using Python to create custom plots of the data, while others used commercial software like Origin or Microsoft Excel. When the dashboard was introduced some users were hesitant to move away from their own personal workflows for comparing data. Introducing the dashboard from the beginning of the collaboration would have mitigated some of these issues. Scientists do not always want to leave the methods they already know due to the inconvenience of changing workflows part way through a project.

**Data longevity and accessibility**

The last lessons learned during this work center on data longevity and accessibility. Data infrastructures require dedicated community efforts and sustained investment to remain operational; otherwise, such tools are at risk of becoming digital ruins [43]. In order to be responsible materials data scientists, application hosting costs and future long-term hosting plans should be discussed early on. To this end, funding and personnel should be allocated to the development and long-term maintenance of a dashboard. While the current initial developments of materials data infrastructures may be purely for the short-term purposes of basic science, funding agencies should prioritize long-term financial support for sustained operation. To move closer to an all-encompassing materials data infrastructure, funding opportunities that bring together materials scientists, computer scientists, and software developers in long-term collaborations is necessary. In the case of a prototype (e.g., this one) where dedicated funding is not present, long-term hosting may not be an option, thus limiting data longevity and accessibility. In our case, we decided to host and maintain the dashboard for the duration of the project and subsequently publish its development as a prototype as opposed to a general-purpose tool. Regardless of whether or not long-term hosting is a possibility, making the core application infrastructure open-source is crucial for enabling knowledge sharing for the next generation of data-sharing apps.

## Future work

In addition to data storage and visualization, we envision two powerful capabilities of future materials dashboards - namely, automated data engineering and integration with ML models to drive autonomous experimental loops. While these have been demonstrated in individual research groups for particular problems [19, 45–48], the generalization of this ability to multi-modal, multi-institutional experimental data in a dashboard is lacking. For example, several data management tools have been developed for either automation of workflows (*e.g.,* HELAO [49], ARES OS [50], MDML [51], ChemOS [52]) or for broad database development (*e.g.*, HTEM [53], MPS [54], MEAD [55]) within a single institution, but not for multi-institutional data management. In our case, the multi-modal, multi-institutional nature of the data required tedious standardization of diverse datasets coming from multiple sources and communication with personnel from each institution, which could not be achieved with any publicly available applications. These difficulties highlight the need for standardized guidelines for data management beyond FAIR [11], as the data management requirements for establishing multi-institutional autonomous loops include more complex data tasks, namely, i) automated data ingestion, engineering, and analysis, ii) seamless communication between institutions and experimental tasks, and iii) the incorporation of domain knowledge While there has been significant progress in the automation of simple workflows, to fully automate materials discovery for complex workflows, standards for the automation of these steps need to be developed (cf. Figure 4 of [48]).

One of the major hindrances to this is the specificity of data engineering for different types of data, again highlighting the difficulty of automating the handling of multi-modal data. Data engineering, which encompasses aggregation, homogenization, featurization, and feature selection, is typically the most time-consuming and critical step for ensuring trustworthy data analysis and ML model training. Due to these challenges, data engineering is often the least automated aspect of data science workflows. However, many of these difficulties can be alleviated through communication with domain experts and the establishment of standards for different types of experiments and materials, as demonstrated in this work for the problem of organizing and aggregating multi-modal, multi-institutional thermoelectric materials data. A schematic of the flow of data into and out of the dashboard for an autonomous loop is shown in **Figure 6**.

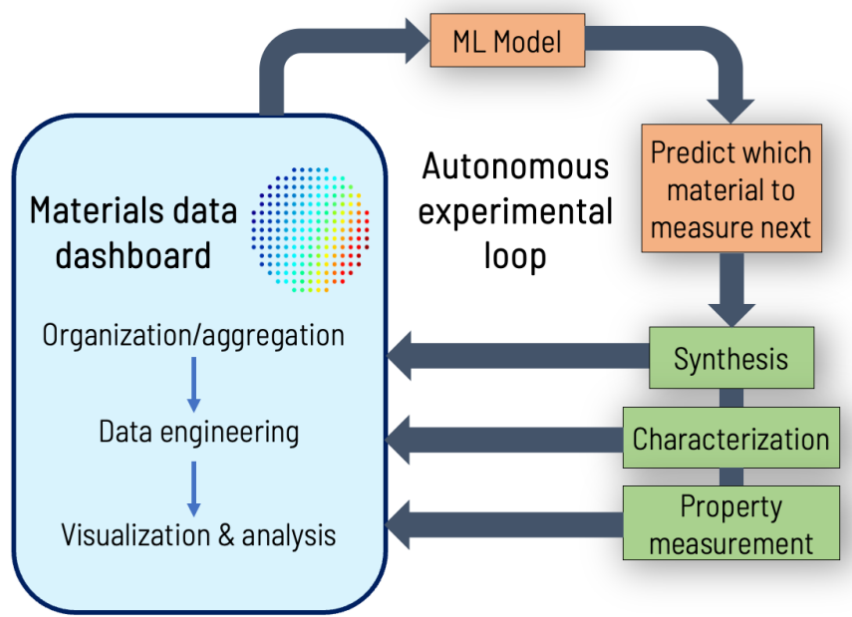

**Figure 6.** Schematic of an autonomous experimental loop with data flowing into and out of a materials dashboard. For a truly autonomous loop, each step would be entirely automated.

From a broader perspective, the long-term vision for materials data infrastructure is an all-encompassing materials dashboard similar to the Materials Data Facility (MDF) [57], where various types of materials data can be automatically uploaded, organized, engineered, and analyzed by different individuals and institutions. This can increase the efficiency of multi-institutional collaborations, where data produced from experiments across the globe can be automatically uploaded to the dashboard from the experimental equipment, eliminating the human middleperson. If data security is required, various levels of permissions can be built into the dashboard. Thus, instead of relying on the collaborators on a single project to produce and transfer data, researchers can search for and use the public data in a web-based dashboard with minimal effort. Not only would this reduce redundancy of experimentation, but it would also enable and encourage replication as well as reduce errors, as "many eyes make any bug shallow". Currently, the MDF allows users to upload individual datasets, but there are no automated tools for combining datasets, running the full data science pipeline, or visualization/analysis. One possible path for achieving this is for a journal or funding agency to collect plotting and analysis scripts for various types of data (e.g., characterization, property measurements) produced by different equipment, each of which is incorporated into the dashboard by publisher or agency developers. For future publications, this can be incentivized or enforced by the publisher or agency, while past publications can be mined by large language models (LLMs) and appropriate prompt engineering (e.g., "Write Python code to reproduce Figure 1"). To maximize the data and code extracted from the literature, full automation of this process would be ideal, which would require an LLM to identify the appropriate prompts from a given paper, another LLM to answer the prompts, and perhaps another model (does not need to be an LLM) to categorize the data and code into material and experiment types. We believe this level of automation is feasible with the recent rate of advancements in LLMs, as long as there are resources dedicated to corpus curation.

However, all of this requires the standardization of *domain-specific ontologies* and of *metadata formatting*, for materials and instruments, within the materials science community, which serve to increase the *interoperability* and *reusability* of data, respectively. These two topics have been increasingly acknowledged as critical to addressing the most pressing materials science problems with data [39, 58]. An *ontology* defines the formal representation of domain knowledge and thus defines what types of data and metadata fully describe a material in a given domain. For example, an ontology for catalyst materials would define what synthesis and processing parameters, instrument metadata, materials properties, device properties/materials, and catalyst performance metrics should be recorded and published as "catalyst materials data". Defining such an ontology, as well as standards regarding data and metadata formatting, could completely alleviate the data standardization issues, and thus the interoperability and reusability issues, in multi-institutional materials data management. These ontologies and metadata format standards then could be published on and enforced by a website like the MDF.

## Conclusion

While computational materials databases such as the Materials Project [15] are extensive, they are limited in the material types and properties that can be robustly modeled, as well as their ability to inform the processing and performance components of PSPP relationships. In order to discover and design transformative materials relevant to energy and climate change, more novel, thoughtful approaches to materials informatics need to be developed, particularly with respect to the management and sharing of experimental data. While there are vast amounts of experimental materials data produced daily, there currently is a lack of centralized, standardized experimental data management. As an initial step toward better data management, here we presented a prototype of a web-based dashboard for exploiting the wealth of information available in a particularly difficult dataset - namely, a multi-modal, multi-institutional combinatorial materials dataset.

In contrast to previous applications for the analysis and visualization of combinatorial data [22, 23], our dashboard i ) can be accessed via a web browser (i.e., is not a desktop application), ii ) integrates directly with Globus [25, 26], a tool already widely used by the scientific research community, and iii ) can aggregate and visualize different data modalities across different wafers and samples. From a broader perspective, our dashboard sets itself from other similar tools [49-55] that have been developed in the high throughput and autonomous materials science communities in its handling of *multi-institutional* data, as all of these tools were designed for data streams within a single institution. Key to the dashboard's development was frequent communication between the developer and the experimental team, which enabled the evolution of the dashboard to better meet the needs of the researchers. Throughout this learning, we developed several recommendations for the development of future materials dashboards, as well as outlined possibilities outside the scope of this work (e.g., integration with ML models). Lastly, in line with the MGI's goals of infrastructure unification and harnessing the power of data, we envision the future development of an all-encompassing materials data management infrastructure based on the prototype developed here, and present strategies to achieve this, particularly through the utilization of LLM models.

# Code availability

The source code for the data portal described in this work is currently available on Bitbucket at the following link: https://bitbucket.org/tecca-data-portal/workspace/snippets/rqnMyX/source-code#file-data-portal-src.zip. Note that this code is made available for reference only, as running the dashboard requires access to a private Globus endpoint.

# Acknowledgements

This work was partially supported by the U.S. Department of Energy, Office of Energy Efficiency and Renewable Energy (EERE), specifically the Advanced Materials & Manufacturing Technologies Office (AMMTO), under contract DE-AC02-76SF00515.

# Conflict of interest

On behalf of all authors, the corresponding author states that there is no conflict of interest.